\documentclass[letterpaper]{article}
\usepackage{authblk}
\usepackage{iccc}
\usepackage{graphicx}
\usepackage{censor}
\usepackage{algpseudocode}
\usepackage{amsmath}
\usepackage{hyperref}
\usepackage{comment}
\usepackage{float}
\usepackage{listings}

\usepackage{times}
\usepackage{helvet}
\usepackage{courier}

\lstnewenvironment{algorithm}[1][] 
{   
    \refstepcounter{nalg} 
    \captionsetup{labelformat=algocaption,labelsep=colon} 
    \lstset{ 
        mathescape=true,
        frame=tB,
        numbers=left, 
        numberstyle=\tiny,
        basicstyle=\scriptsize, 
        keywordstyle=\color{black}\bfseries\em,
        keywords={,input, output, return, datatype, function, in, if, else, foreach, while, begin, end, } 
        numbers=left,
        xleftmargin=.04\textwidth,
        #1 
    }
}
{}

\title{SYMPLEX: Controllable Symbolic Music Generation using Simplex Diffusion with Vocabulary Priors}

\author[]{Nicolas Jonason}
\author[]{Luca Casini}
\author[]{Bob L.T. Sturm}
\affil[]{KTH Royal Institute of Technology, Stockholm, Sweden}

\begin{document} 

\maketitle
\begin{abstract}
\begin{quote}
We present a new approach for fast and controllable generation of symbolic music
based on the simplex diffusion, which is essentially a diffusion process operating on probabilities
rather than the signal space.
This objective has been applied in domains such as natural language processing
but here we apply it to generating 4-bar multi-instrument music loops using 
an orderless representation. 
We show that our model can be steered with vocabulary priors, which affords a considerable level control over the music generation process, 
for instance, infilling in time and pitch and choice of instrumentation --- all without task-specific model adaptation or applying extrinsic control.

\end{quote}
\end{abstract}

\section{Introduction}
Diffusion models are a class of machine learning models that have been applied to generating images \cite{stable}, audio \cite{crash}, video \cite{bartal2024lumiere}, and natural language \cite{han2023ssdlm} and more.
These models define a forward diffusion process, e.g., adding noise to an image, and train a neural network to reverse this process.
Advantages of diffusion models over auto-regressive models include the ease of plug-and-play guidance using external models as well as the ability to trade-off inference speed and quality after training.

For symbolic music generation, at least three types of diffusion models appear in the litterature. The first type 
uses discrete diffusion operating on the signal domain, which involves defining transition probabilities between states \cite{ijcai2023p648}.
The second type defines the diffusion process in embedding space, rendering it continuous \cite{stanford}. In some variants, the diffusion process operates in a lower-dimensional latent space, which allows for faster training and inference \cite{mittal}. Finally, a third approach represents the music as a piano-roll image and treats the pixel values as continuous \cite{polyfusion,huang2024symbolic}.

In this short paper, we explore a different type of diffusion: \textit{Simplex Diffusion} (SD) \cite{richemond2022categorical,han2023ssdlm,floto2023diffusion,mahabadi2023tess}.
This applies the diffusion process on probability distributions over the signal domain 
rather than on the signal itself, rendering the diffusion process continuous even if the signal is discrete. 
Han et al. (\citeyear{han2023ssdlm}) argue that SD has an advantage in its plug-and-play-guidance ability over diffusion models whose diffusion process operates directly in embedding space. Namely, SD requires only that the external guidance model shares the same vocabulary whereas the other approach require that the external guidance models are developed from scratch. In addition to the ease of external control, we argue that SD models offer additional advantages in terms of internal control. Since the diffusion process operates on the signal's probability distribution, we can easily steer the generation by injecting priors on the probability distributions during inference.
We demonstrate this by introducing \emph{SYMPLEX}, a simplex diffusion model operating on an orderless note-set representation of 4-bar multi-instrument MIDI loops. 
SYMPLEX is capable of infilling in time and pitch, conditioning on instrumentation, tonality and rhythm, generating variations, and more without task-specific fine-tuning or extrinsic control.

We summarize our contributions as follows:
1) We present SYMPLEX, a simplex diffusion model operating on an orderless note-set representation of 4-bar multi-instrument MIDI loops. 
To our knowledge this is the first use of simplex diffusion for symbolic music generation;
 2) We show how a simplex diffusion model can be steered with vocabulary priors to address a range of different music generation tasks;
3) We adapt a context-based loop extraction technique from \cite{adkins2023loopergp} to MIDI and extend it with a metrical-structure heuristic to obtain better loops.

We provide a demo website with examples showcasing our model successes and failures across multiple tasks.
\footnote{
    \href{https://erl-j.github.io/slm-demo/}{https://erl-j.github.io/slm-demo/}
}

\section{Simplex Diffusion for Controllable Symbolic Music Generation}

While many variants of simplex diffusion have been proposed, our work is based on SSD-LM \cite{han2023ssdlm}. 
SSD-LM generates natural language sequences of arbitrary length by windowed simplex diffusion based on preceding context. 
Additionally, they also explore extrinsic control with classifier guidance. 
For simplicity, we scope of our work to fixed length sets and focus only on control with vocabulary priors.
Another difference with our work is that while SSD-LM deals with \textit{sequences}, we instead operate on \textit{sets} of note-events, each consisting of 9 attributes. While this makes no difference in terms of the general framework, this does lead to some differences in the neural network architecture (see the Architecture subsection below).
While our work is primarily focused on the application of SD to symbolic music, we also make a minor contribution to the SD framework by demonstrating that we can easily guide the generation with vocabulary priors.

We will now go over the important parts of the simplex diffusion framework. 
For a more thorough introduction to the SD framework, we refer to \cite{han2023ssdlm}.
Note that we adjusted the SSD-LM framework in order to more easily apply the vocabulary priors.
In SSD-LM, the neural network takes logits as input, whereas in our work, the neural network instead takes probabilities as input.
\paragraph{Training process}
We train a neural network $\theta$ to recover data samples from noisy probabilities.
Each training step starts by drawing a sample $w$ from the training dataset that is then mapped to logits of scale $K$ using the following $\textit{logit-generation}$ operation:
\begin{equation}
\tilde{w}_{0_{(i)}} = 
\begin{cases} 
+K & \text{when } w = V_{(i)} \\
-K & \text{when } w \neq V_{(i)}
\end{cases}
\end{equation}
We then draw the time $t\sim\mathcal{U}_{[0,1]}$ which determines the degree of noise $\alpha_t$ using a predetermined noise schedule.
We then obtain our noisy logits $\tilde{w}_t$ by applying Gaussian noise:
\begin{equation}
\tilde{w}_t = \tilde{w}_0\sqrt{\alpha_t} + \epsilon\sqrt{1-\alpha_t}, \text{ where } \epsilon \sim \mathcal{N}(0,K^2\mathbf{I})
\end{equation}
We then apply apply a softmax to the noisy logits to obtain probabilities $p_t=\text{softmax}(\tilde{w}_t)$.

The probabilities are then passed to a neural network $\theta$ together with the time $t$ which returns denoised logits. We then update the network according to the cross entropy loss:
\begin{equation}
\mathcal{L}_\theta =-\sum_{j=0}^{|w|}p_\theta(w^j|p_t,t) 
\end{equation}
\paragraph{Inference}
To generate a new sample, we start from randomly initialised probabilities and iteratively refine these across $T$ steps.
The probabilities are initialised as follows:
\begin{align}
        \tilde{w}_T \sim \mathcal{N}(0,K^2,I) \\
    p_T=\text{softmax}(\tilde{w}_T)
\end{align}
Then, we iteratively produce denoised logits with the neural network $\theta$ which we then use to produce the probabilities for the next iteration. 
Iterating from step $s=T$ to $1$:
\begin{align}
    t &= \frac{\text{s}}{T} \\
    \hat{\tilde{w}}_0 &= \theta(p_t,t) \label{eq:t}\\
    p_0 &= \text{softmax}(\hat{\tilde{w}}_0) \\
    \hat{w} &= \text{sample}(p_0) \\
    {\hat{\tilde{w}}_0}'&= \text{logit-generation}(\hat{w}) \\
    \tilde{w}_{t-1} &= {\hat{\tilde{w}}_0}'\sqrt{\alpha_{t-1}} + \epsilon_t\sqrt{1-\alpha_{t-1}} \\
    p_{t-1} &=\text{softmax}(\tilde{w}_{t-1})
\end{align}

where $\epsilon_t \sim \mathcal{N}(0,K^2\mathbf{I})$.
At the final iteration we return $\hat{w}$, which represents our generated sample.
Note how the procedure for producing the noisy logits for the subsequent step involves sampling. 
The SSD-LM authors propose multiple approaches for this sampling step. This work uses their top-p sampling approach.
\paragraph{Controlling generation with vocabulary priors}
\begin{figure*}[ht]
\centering
    \includegraphics[trim={0 0 0 0},width=0.97\linewidth]{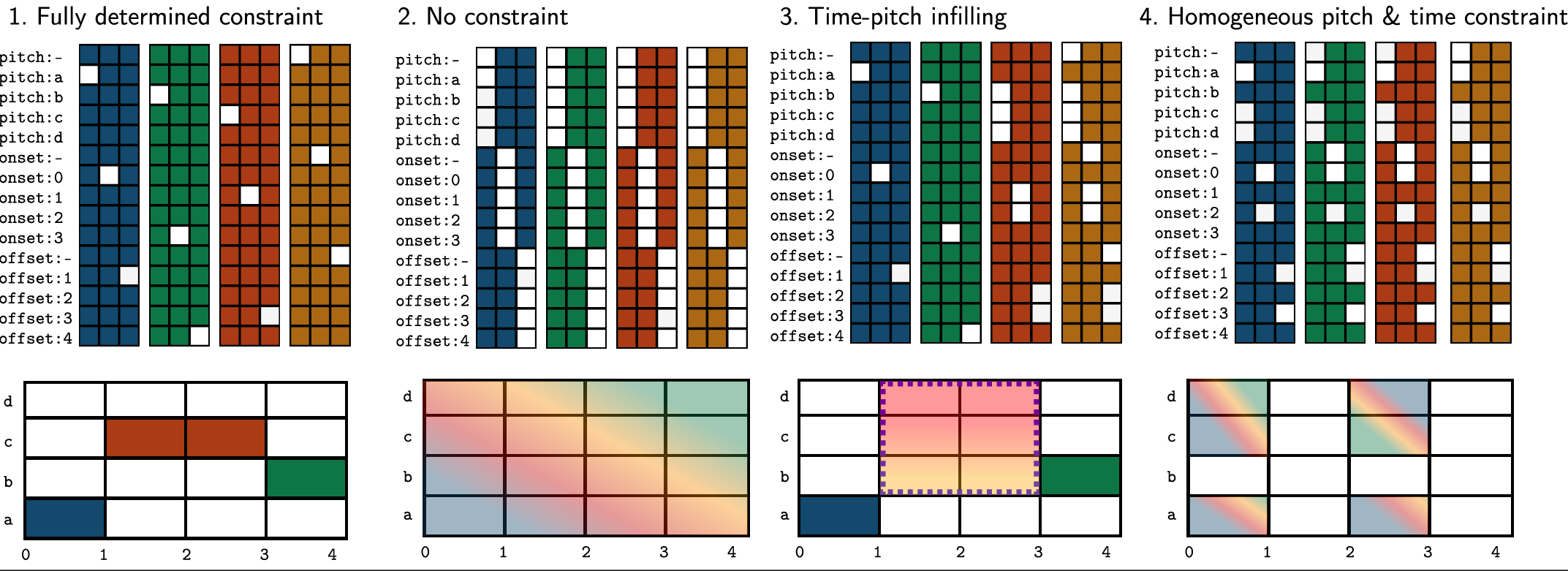}
    \caption{
    \textbf{Expressing various symbolic music generation tasks as priors on unordered representations.}
    For clarity, this figure uses a toy representation of music with 4 pitches, 
    4 discrete onset times 
    and 4 discrete offset times. 
    The upper row shows vocabulary priors where non-zero probabilities are represented with white cells.
    The bottom row illustrates the constraints in piano roll form.
    Each note event, colour coded in the figure, has three attribute columns representing the pitch, onset and offset constraints respectively.
    A colour gradient in the piano roll indicates that the note event(s) of the corresponding hue might be generated in the region.
    \textbf{1.} shows a fully determined musical piece containing 3 notes. Notice how the orange note has all attributes set to undefined,
    indicating an inactive note.
    \textbf{2.} shows a completely uninformative prior. 
    Notice how the pitch, onset and offset vocabularies don't overlap.
    \textbf{3.} shows a prior representing a time-pitch infilling task with the piece depicted in \textbf{1.} as the input. 
    Notice how the red note and orange notes differ in their columns. 
    Unlike red which is guaranteed to be active,
    orange on the other hand might be active or not.
    This allows us to express precise ranges on the number of notes we want. 
    In this case, we are saying: "infill this region with at least one note".
    \textbf{4.} shows how we can use priors to control tonality and rhythm.
    }
\label{fig:masks}
\end{figure*}
The SSD-LM authors experiment with using external attribute classifiers to control the generation by nudging the logits produced by $\theta$ in the direction of the classifiers gradient. 
We explore a different type of control, which does not require external models but rather steers generation using vocabulary priors on the input elements. 
The motivation for this is that vocabulary priors on the input elements allows us to express a lot of different generation tasks as seen in Figure \ref{fig:masks}.
It should be noted that our approach does not preclude the use of plug-and-play guidance with external models.
Future work will explore combining these two control approaches.

We inject the vocabulary prior $p_v$ by multiplying it with $p_t$ and re-normalising before applying the neural network $\theta$. 
We replace equation \ref{eq:t} with:
\begin{align}
\hat{\tilde{w}}_0=\theta(\text{normalize}(p_t \times p_v),t) 
\end{align}

We also an additional safeguard to prevent the model from generating samples which violate the prior by setting probabilities to zero if the prior has probability zero, and probability one if the prior has probability one.

\section{Demonstration}\label{sec:exp}
In this section we describe the loop dataset, representation, model architecture, training as well as the tasks used to demonstrate SYMPLEX.
\subsection{Building A MIDI Loop Dataset}\label{sec:dataset}
We set our experiments in the domain of 4-bar multi-instrument MIDI loops extracted from the MetaMIDI Dataset \cite{metamidi}.
The dataset contains 430k multi-track MIDI files of popular songs from a variety of styles. 
We make the following considerations for splitting the dataset and extracting loops.

\subsubsection{Splitting the dataset}\label{sec:split}
10\% of the data is held out to be used as a test set. 
The remainder is further split into two portions: 90\% for training and 10\% for evaluation.
Although the MetaMIDI dataset does not contain exact duplicates of the same MIDI files, multiple MIDI files can correspond the same song and vice-versa. To mitigate the chance of two or more MIDI files representing the same song ending up in different splits we construct a bipartite graph with the MIDI file-hashes and the Spotify track ids.
We then split the data according to connected components of this graph.
The MIDI files which have not been matched to Spotify tracks ids are assigned randomly to each split. 
We perform the split before the loop extraction, so that loops from the same MIDI file will not appear in different splits.
\subsubsection{Extracting Loops}\label{sec:loops}
At least two approaches have been proposed for extracting loops from symbolic music. \cite{korea} leverages an external dataset of audio loops and then uses cross-modal anomaly detection to detect whether a MIDI section is loopable according to it's musical content. \cite{adkins2023loopergp} instead uses a context based approach, and propose an efficient algorithm to detect sections bookended by a musical phrase in the domain of Guitar Pro tablature files. We adapt the latter approach to MIDI and extend it by adding additional requirements on the metrical structure. The reason for including metrical structure cues as an additional heuristic is that we find that the book-ended-phrase-heuristic often yields loops where the first beat does not align with the perceived first beat.
The extended book-ended-phrase-heuristic works as follows:
\textit{A section is considered a loop if the following is true:}
1) 
\textit{it is book-ended by a musical phrase};
2) \textit{its start coincides with an important metrical boundary};
and 3) \textit{its start is the most important metrical boundary of the section}.

To detect the metrical structure, we use a pre-trained hierarchical metrical structure analysis model proposed by Jiang and Xia (\citeyear{sshmsm}). The model outputs the probabilities of metrical structure boundaries at each beat of the piece with an 8-level hierarchy. We first pre-filter the dataset, keeping files which have a only one tempo and a 4/4 time signature, as this is the only time-signature supported by the structure analysis.
Using the metrical structure analysis, we crop the piece to start at the the first detected downbeat (A level-4 boundary probability larger than 0.5). For each cropped frame, we then extract piano rolls with a $1/8^{th}$ note time resolution and 139 channels. The first 11 channels represent the combination of the pitch classes for every instrument in the MIDI file, irrespective of octave. The remaining 128 channels are instead used to represent all possible drum sounds. We then find loop candidates by looking for book-ended sections using a bookend length threshold of 2 frames -- i.e a quarter note. We also filter the candidates to only keep loops that are 4-bars long.
Using the output from the metrical structure analysis model, we only keep candidates whose start coincide with a level-5 boundary and where no metrical boundary of level higher than 5 occurs after start. 
The final loop dataset consists of $\sim 250,000$ 4-bar MIDI loops.

\subsection{MIDI Loop Representation}

The choice of representation has direct implications on the ease with which we can later implement different types of control that we can exert on the system.
Ordered representations of music bind information about the notes such as time, instrument and pitch to their absolute or relative position within the sequence.
With this in mind and following \cite{wang2021musebert}, we use an \emph{unordered set} representation as it gives us the freedom to regenerate any note attribute without concern of violating the representation's syntax. We represent a MIDI loop as a set of note event tuples. Each note event tuple contains 9 attributes conveying its instrument, pitch, onset and offset, velocity, as well as the loops global information loop tempo and loop style. We represent \emph{inactive note events} by setting its list of attributes to their respective undefined tokens.

\begin{itemize}

\item\textbf{Instrument:} We use 17 tokens for general MIDI instrument classes (Piano, Bass, Flute etc...), a special instrument token for drums and one ``undefined instrument''.
\item\textbf{Pitch:} 127 tokens for pitch, 47 tokens for drum pitch and one token for ``undefined pitch''.
\item\textbf{Onset beat:} 16 tokens for beats 0 to 16, one token for ``undefined onset beat''.
\item\textbf{Onset tick:} 24 tokens for ticks 0 to 24, one token for ``undefined offset beat''.
\item\textbf{Offset beat:} 16 tokens for beats 0 to 16, one special token for drums offset for which we do not consider offsets, which are ignored one token for ``undefined offset beat''.
\item\textbf{Offset tick:} 24 tokens for beat and one token for drums for which we do not consider offsets and one token for ``undefined offset tick''.
\item\textbf{Velocity:} 32 tokens for velocity bins and one token for ``undefined velocity''.
\item\textbf{Tempo:} 16 tokens for tempo bins spanning 50-200 bpm and one token for ``undefined tempo''.
\item\textbf{Tag:} We use 39 different tags, corresponding to the annotations in the MetaMIDI dataset. 
If a loop is associated with multiple genres, we assign a random genre during training. 
We also use a special token ``other'' for loops with no genre annotation and an ``undefined tag'' token.
\end{itemize}

\subsection{Architecture and training details}
The role of the neural network $\theta$ is to predict the entire input from the probability distributions concerning each element in the input as well as a time variable $t$. We use a BERT-like transformer encoder-only architecture \cite{bert}.
With our representation, a loop is represented as a set of $N$ note event tuples each consisting of $A$ attributes, totalling $N\times A$ input elements. The input to the network will therefore consist of $N \times A$ probability distributions. In order to embed a probability distribution, SSD-LM embeds all tokens in the vocabulary and then sum the token embeddings weighted by the probability of each token. However, recall that our representation involves $A$ attributes with distinct sub-vocabularies $V_a\in V$. We can avoid the neural network having to the learn the syntax by multiplying the probabilities with a syntax prior and then re-normalizing prior to taking the weighted sum.
Recall that the neural networks input has dimensionality $N\times A$ where $N$ is the number of note events and $A$ is the number of attributes per note event. While it is possible to apply the transformer encoder to the entire set of $N\times A$ elements, the self-attention operation inside the transformer would have complexity $O((N\times A)^2)$. We can reduce the complexity of the self-attention to $O(N^2)$ by aggregating attribute embeddings of each note with a summing operation \cite{wang2021musebert}.
Similarly to how we enforce the syntax at the embedding step, we recover attribute specific logits from the transformer output by setting the logits to $-\infty$ if the corresponding token is not in the target attribute's sub-vocabulary.

The transformer encoder has 8 layers, 8 heads, hidden size 512, feedforward size 1024.
We use ADAM with a base learning rate of $10^{-3}$, epoch-wise learning rate decay of $0.99$ and batch size $200$. 
We train for $\sim100$ hours on a single RTX 3090.

\subsection{Generation tasks}

\begin{figure}
    \centering
    \includegraphics[width=0.8\columnwidth]{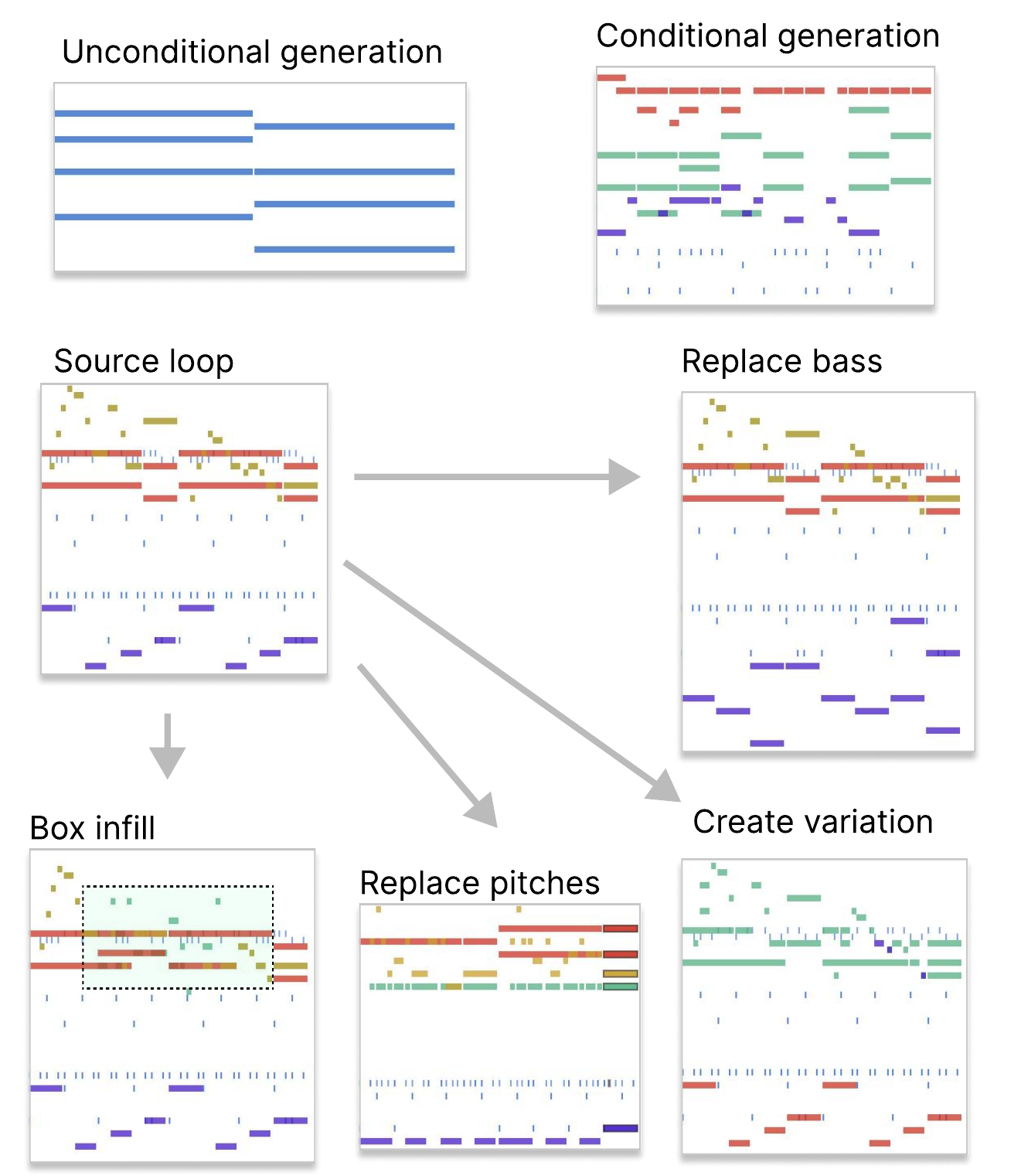}
    \caption{Examples of the tasks we experimented with. More examples with audio are on the website.}
    \label{fig:tasks}
\end{figure}

We demonstrate SYMPLEX on multiple tasks: 
unconditional generation, generation with a instrumentation/pitch constraint, as well as several editing tasks. 
As seen in Figure \ref{fig:tasks}, the editing tasks include box infilling, generating variations of a loop, regenerating bass, replacing pitches and more.
For the editing tasks, we selected 6 loops from the test data with a variety of styles and instrumentation, to use as our source. 
For each task, we generate priors using task-specific scripts which takes source loop and task specific parameters. 
For each task we also adjusted the top-p threshold as well as $T$ (between $50$ and $300$ steps). Note that we did not adjust these parameters for each source loop individually.
More details and audio examples are available at \href{https://erl-j.github.io/slm-demo/}{https://erl-j.github.io/slm-demo/}.

\section{Future work}
While our model is capable of producing plausible outputs in the different generation tasks, each task requires tuning of the number of steps $T$ and top-p. While we did not experiment with adjusting $T$ and top-p by source loop, we suspect that the musical content of the source loop also plays a role and needs to be considered in setting the parameters.
While it is commonplace for diffusion models to require that users adjust parameters settings according to different generation scenarios, it would improve the workflow considerably if this parameter tweaking could be avoided. This will be the main direction for our future work.

\bibliographystyle{iccc}
\bibliography{iccc}

\end{document}